\documentclass[journal]{IEEEtran}
\usepackage[dvips]{graphicx}
\def\beginofnote{\indent {\it Note: }}
\def\beginofproof{\indent {\it Proof: }}
\def\endofproof{\hfill\rule{6pt}{6pt}}
\newtheorem{theorem}{Theorem}
\newtheorem{lemma}{Lemma}

\newtheorem{example}{Example}

\begin{document}

\title{Results on Parity-Check Matrices with Optimal Stopping
and/or Dead-End Set Enumerators}

%\author{Khaled A.S. Abdel-Ghaffar,~\IEEEmembership{Member,~IEEE,}
%        and Jos H. Weber,~\IEEEmembership{Senior Member,~IEEE}% <-this % stops a space
\author{Jos H. Weber,~\IEEEmembership{Senior Member,~IEEE,}
        and Khaled A.S. Abdel-Ghaffar,~\IEEEmembership{Member,~IEEE}% <-this % stops a space
\thanks{Manuscript submitted to IEEE Transactions on Information Theory, July 7, 2006.
        The authors were supported by the Dutch Ministry of
Economic Affairs through the Airlink research project DTC.5961, and
by the NSF through
grants CCR-0117891 and ECS-0121469. This paper was presented in part
at the International Symposium on Turbo Codes, Munich,
Germany, April 3--7, 2006.}% <-this % stops a space
\thanks{Jos H. Weber is with the Delft University
of Technology, Faculty of EEMCS, Mekelweg 4, 2628 CD Delft, The
Netherlands, j.h.weber@tudelft.nl.
Khaled A.S. Abdel-Ghaffar is with the University of
California, Department of ECE, Davis, CA 95616,USA,
ghaffar@ece.ucdavis.edu.}}

\markboth{Submission to IEEE Transactions on Information Theory,
July 7, 2006. } {
%Shell \MakeLowercase{\textit{et al.}}: Bare Demo of IEEEtran.cls for Journals
}

\maketitle

\begin{abstract}
The performance of iterative decoding techniques for linear block
codes correcting erasures depends very much on the sizes of the
stopping sets associated with the underlying Tanner graph, or,
equivalently, the parity-check matrix representing the code. In
this paper, we introduce the notion of dead-end sets to explicitly
demonstrate this dependency. The choice of the parity-check matrix
entails a trade-off between performance and complexity. We give
bounds on the complexity of iterative decoders achieving optimal
performance in terms of the sizes of the underlying parity-check
matrices. Further, we fully characterize codes for which the
optimal stopping set enumerator equals the weight enumerator.
\end{abstract}


\IEEEpeerreviewmaketitle

\section{Introduction} \label{intro}

%\subsection{Subsection Heading Here}
%Subsection text here.

% needed in second column of first page if using \pubid
%\pubidadjcol

%\subsubsection{Subsubsection Heading Here}
%Subsubsection text here.

\PARstart{I}{terative} decoding techniques, especially when
applied to low-density parity-check (LDPC) codes, have attracted a
great attention recently. In these techniques, decoding is based
on a Tanner graph determined by a parity-check matrix of the code,
which does not necessarily, and typically does not, have full
rank. It is well known that the performance of iterative decoding
algorithms in case of binary erasure channels depends on the sizes
of the stopping sets associated with the Tanner graph representing
the code \cite{DPTRU}. Several interesting results on stopping
sets associated with Tanner graphs of given girths are given in
\cite{KSXR,OUVZ}. There are more specific results for classes of
codes represented by particular Tanner graphs, see, e.g.,
\cite{KV,AW,XF}, as well as more general results pertaining to
ensembles of LDPC codes, see e.g., \cite{BM,DPTRU,IKSS,OKVZ}.

In this paper, we define the notion of dead-end sets to explicitly
show the dependency of the performance on the stopping sets. We
then present several results that show how the choice of the
parity-check matrix of the code, which determines decoding
complexity, affects the stopping and the dead-end sets, which
determine decoding performance. Our study differs from the
aforementioned studies, but agrees with the studies by Schwartz
and Vardy \cite{SV}, Hollmann and Tolhuizen \cite{HT},
Han and Siegel \cite{HS}, and Weber
and Abdel-Ghaffar \cite{WA}, in its focus on the relationship
between the stopping sets on one hand and the underlying code
representation, rather than the code itself, on the other hand.
Since linear algebra is used to study this relationship, for our
purpose, parity-check matrices are more convenient than the
equivalent Tanner graphs for code representation.

Let $\cal C$ be a binary linear $[n,k,d]$ block code, where $n$,
$k$, and $d$ denote the code's length, dimension, and Hamming
distance, respectively. Such a code is a $k$-dimensional subspace
of the space of binary vectors of length $n$, in which any two
different elements differ in at least $d$ positions. The set of
codewords of $\cal C$ can be defined as the null space of the row
space of an $r\times n$ binary parity-check matrix ${\bf
H}=(h_{i,j})$ of rank $n-k$. Assuming all rows in $\bf H$ are
different,
\begin{equation}
n-k\le r\le 2^{n-k}.
\end{equation}
The row space of $\bf H$ is the $[n,n-k,d^\perp]$ dual code ${\cal
C}^{\perp}$ of $\cal C$.

The support of a binary word ${\bf x}=(x_1,x_2,\ldots,x_n)$ is the
set $\{j:x_j\neq 0\}$ and the weight of ${\bf x}$ is the size of
its support. For the zero word ${\bf 0}=(0,0,\ldots,0)$, the
support is the empty set, $\emptyset$, and the weight is zero.
Since a binary word $\bf x$ is a codeword of ${\cal C}$ if and
only if ${\bf x}{\bf H}^{\rm T}={\bf 0}$, the parity-check matrix
$\bf H$ gives rise to $r$ parity-check equations, denoted by
\begin{equation} \label{PCE}
\mbox{${\rm PCE}_i({\bf x}):$ $\sum_{j=1}^n h_{i,j}x_j=0$ for
$i=1,2,\ldots,r$.}
\end{equation}
An equation ${\rm PCE}_i({\bf x})$ is said to check $\bf x$ in
position $j$ if and only if $h_{i,j}=1$.

On the binary erasure channel, each bit of the transmitted
codeword is erased with probability $\epsilon$, while it is
received correctly with probability $1-\epsilon$, where
$0<\epsilon<1$. For a received word ${\bf
r}=(r_1,r_2,\ldots,r_n)$, the erasure set is
\begin{equation}
{\cal E}_{\bf r}= \{j:r_j\neq 0,1\}.
\end{equation}
A received word can be decoded unambiguously if and only if it
matches exactly one codeword of $\cal C$ on all its non-erased
positions. Since $\cal C$ is a linear code, this is equivalent to
the condition that the erasure set ${\cal E}_{\bf r}$ does not
contain the support of a non-zero codeword. If ${\cal E}_{\bf r}$
does contain the support of a non-zero codeword, then it is said
to be {\em incorrigible}. A decoder for $\cal C$ which achieves
unambiguous decoding whenever the erased set is not incorrigible
is said to be optimal for the binary erasure channel. An
exhaustive decoder searching the complete set of codewords is
optimal. However, such a decoder usually has a prohibitively high
complexity.

Iterative decoding procedures may form a good alternative,
achieving close to optimal performance at much lower complexity
\cite{LMSS},
in particular for LDPC codes. Here, we consider a well-known
algorithm, often expressed in terms of a Tanner graph, which
exploits the parity-check equations in order to determine the
transmitted codeword. Initially, we set ${\bf c}={\bf r}$ and
${\cal E}_{\bf c}={\cal E}_{\bf r}$. If ${\rm PCE}_i({\bf c})$
checks $\bf c$ in exactly one erased position $j^\ast$, then we
use (\ref{PCE}) to set
\begin{equation} \label{cj*}
c_{j^\ast}=\sum_{j\not\in{\cal E}_{\bf c}} h_{i,j}c_j
\end{equation}
and we remove $j^\ast$ from the erasure set ${\cal E}_{\bf c}$.
Applying this procedure iteratively, the algorithm terminates if
there is no parity-check equation left which checks exactly one
erased symbol. Erasure sets for which this is the case have been
named {\em stopping sets} \cite{DPTRU}. In case the final erasure
set ${\cal E}_{\bf c}$ is empty, the iterative algorithm retrieves
all erased symbols, and thus the final word $\bf c$ is the
transmitted codeword. In case the final erasure set ${\cal E}_{\bf
c}$ is a non-empty stopping set, the iterative decoding process is
unsuccessful. The final erasure set ${\cal E}_{\bf c}$ is the
union of the stopping sets contained in ${\cal E}_{\bf r}$, and
thus ${\cal E}_{\bf c}$ is empty if and only if ${\cal E}_{\bf r}$
contains no non-empty stopping set. Therefore, we introduce the
notion of a {\em dead-end set} for an erasure set which contains
at least one non-empty stopping set. In summary, on the binary
erasure channel, an optimal decoder is unsuccessful if and only if
${\cal E}_{\bf r}$ is an incorrigible set, and an iterative
decoder is unsuccessful if and only if ${\cal E}_{\bf r}$ is a
dead-end set.

This paper is organized as follows. In Section~\ref{definition} we
characterize codeword supports, incorrigible sets, stopping sets,
and dead-end sets in terms of a parity-check matrix and derive
basic results from this characterization. We also review results
from \cite{SV} and \cite{HT} which are most relevant to this work.
Dead-end sets and stopping sets are studied in Sections~\ref{sec
ds} and \ref{sec ss}, respectively. Conclusions are presented in
Section~\ref{Conc}.

\section{Definitions and Preliminaries} \label{definition}

Again, let $\cal C$ be a linear binary $[n,k,d]$ block code with
an $r\times n$ binary parity-check matrix ${\bf H}=(h_{i,j})$ of
rank $n-k$. Let ${\cal S}$ be a subset of $\{1,2,\ldots,n\}$. For
any $\bf H$ and $\cal S$, let ${\bf H}_{\cal S}$ denote the
$r\times |{\cal S}|$ submatrix of ${\bf H}$ consisting of the
columns indexed by ${\cal S}$. A set $\cal S$ is the support of a
codeword if and only if all rows in ${\bf H}_{\cal S}$ have even
weight, i.e., if and only if
\begin{equation} \label{eqCW}
|\{j\in {\cal S}: h_{i,j}=1\}|\equiv 0 (2) \hspace{0.5cm}\forall
i=1,2,\ldots,r.
\end{equation}
A set $\cal S$ is an incorrigible set if and only if it contains
the support of a non-zero codeword. A set $\cal S$ is a stopping
set for the parity-check matrix $\bf H$ if and only if ${\bf
H}_{\cal S}$ does not contain a row of weight one, i.e., if and
only if
\begin{equation} \label{eqSS}
|\{j\in {\cal S}: h_{i,j}=1\}|\neq 1 \hspace{0.5cm}\forall
i=1,2,\ldots,r.
\end{equation}
Hence, the support of any codeword is a stopping set. A set $\cal
S$ is a dead-end set for the parity-check matrix $\bf H$ if and
only if it contains a non-empty stopping set.

The polynomial $A(x)=\sum_{i=0}^n A_i x^i$, where $A_i$ is the
number of codewords of weight $i$, is called the {\em weight
enumerator} of code $\cal C$. Similarly, $I(x)=\sum_{i=0}^n I_i
x^i$, where $I_i$ is the number of incorrigible sets of size $i$,
is called the {\em incorrigible set enumerator} of $\cal C$.
Clearly,
\begin{equation}
d  =  \min\{i\ge 1:A_i> 0\} = \min\{i\ge 0:I_i> 0\}
\end{equation}
and
\begin{equation} \label{basicA}
A_i =\left\{\begin{array}{cl} 1 & \mbox{if $i=0$,}  \\
0 & \mbox{if $1\le i\le d-1$.}  \end{array}\right.
\end{equation}
The incorrigible set enumerator satisfies
\begin{equation} \label{basicI}
I_i =\left\{\begin{array}{cl}  0 & \mbox{if $0\le i\le d-1$,}  \\
A_i &\mbox{if $i=d$,} \\
{n \choose i} & \mbox{if $n-k+1\le i\le n$,} \\
\end{array}\right.
\end{equation}
where the last property follows from the observation that any set
$\cal S$ of size $|{\cal S}|>n-k$ contains the support of a
non-zero codeword as the rank of ${\bf H}_{\cal S}$ is at most
$n-k$.

The polynomials $S(x)=\sum_{i=0}^n S_i x^i$, where $S_i$ is the
number of stopping sets of size $i$, and $D(x)=\sum_{i=0}^n D_i
x^i$, where $D_i$ is the number of dead-end sets of size $i$, are
called the {\em stopping set enumerator} and the {\em dead-end set
enumerator}, respectively, of parity-check matrix $\bf H$. From
the observation that (\ref{eqCW}) and (\ref{eqSS}) are equivalent
for sets $\cal S$ with $|{\cal S}|\le 2$, it follows that
\begin{equation} \label{012}
S_i=A_i \mbox{ and } D_i=I_i \mbox{ if } 0\le i\le 2.
\end{equation}
In particular, $S_0=1$, $S_1=S_2=0$, and $D_0=D_1=D_2=0$ for any
parity-check matrix of a code of minimum distance $d\ge 3$.

Let $s$ denote the smallest size of a non-empty stopping set (and
thus the smallest size of a dead-end set), i.e.,
\begin{equation} \label{defs}
s  =  \min\{i\ge 1:S_i> 0\}=\min\{i\ge 0:D_i> 0\}.
\end{equation}
The number $s$ is called the stopping distance for the
parity-check matrix $\bf H$ in \cite{SV}.  For any parity-check
matrix $\bf H$ of a binary linear $[n,k,d]$ block code $\cal C$,
it holds that the stopping set enumerator satisfies
\begin{equation} \label{basicS}
S_i =\left\{\begin{array}{cl}  1 & \mbox{if $i=0$,}  \\
0 & \mbox{if $1\le i\le s-1$,}  \\
{n \choose i} & \mbox{if $n-d^\perp+2\le i\le n$.} \\
\end{array}\right.
\end{equation}
where the first property follows from (\ref{012}) and
(\ref{basicA}), the second property follows from the definition of
$s$, and the third property follows from the fact that the weight
of any row in ${\bf H}_{\cal S}$ is either $0$ or at least equal
to $d^\perp-(d^\perp-2)=2$ for any $\cal S$ with $|{\cal S}|\ge
n-d^\perp+2$. Further, again for any parity-check matrix $\bf H$,
it follows from the definitions of the various enumerators,
(\ref{basicI}), and (\ref{basicS}), that the dead-end set
enumerator satisfies
\begin{equation} \label{basicD}
D_i =\left\{\begin{array}{cl}
0 &\mbox{if $0\le i\le s-1$,} \\
S_i & \mbox{if $i=s$,}  \\
{n \choose i} & \mbox{if $n-k+1\le i\le n$.} \\
\end{array}\right.
\end{equation}

For code ${\cal C}$ on the binary erasure channel, the probability
of unsuccessful decoding (UD) for an optimal (OPT) decoder is
\begin{equation} \label{POPT}
P^{\rm OPT}_{\rm UD}({\cal C}) = \sum_{i=d}^n
I_i\epsilon^i(1-\epsilon)^{n-i}\sim I_d\epsilon^d=A_d\epsilon^d.
\end{equation}
Similarly, the probability of unsuccessful decoding  for an
iterative (IT) decoder based on parity-check matrix $\bf H$ is
\begin{equation} \label{PIT}
P^{\rm IT}_{\rm UD}({\bf H}) = \sum_{i=s}^n
D_i\epsilon^i(1-\epsilon)^{n-i}\sim D_s\epsilon^s=S_s\epsilon^s.
\end{equation}
Hence, these two probabilities are completely determined by the
incorrigible and dead-end set enumerators. Notice from
(\ref{POPT}) and (\ref{PIT}) that iterative decoding is optimal if
and only if $D(x)=I(x)$. At small erasure probabilities, $P^{\rm
OPT}_{\rm UD}({\cal C})$ and $P^{\rm IT}_{\rm UD}({\bf H})$ are
dominated by the terms $A_d\epsilon^d$ and $S_s\epsilon^s$,
respectively. Actually, for sufficiently small values of
$\epsilon$, the parameters $d$ and $s$ are the most important
parameters characterizing the performance of optimal decoding and
iterative decoding, respectively. In (\ref{012}) it is stated that
if $i\le 2$, then $S_i=A_i$. Therefore, $s=d$ for any parity-check
matrix $\bf H$ of a code with $d\le 3$, which is derived as
Theorem~3 in \cite{SV}. Here, we show that this cannot be extended
further.

\begin{theorem} \label{badH}
For any code $\cal C$ with Hamming distance $d\ge 4$, there exists
a parity-check matrix $\bf H$ for which $s=3$.
\end{theorem}

\beginofproof
We may order the positions so that $\cal C$ has a codeword
composed of $d$ ones followed by $n-d$ zeros. In particular, the
first $d$ columns in any given parity-check matrix of $\cal C$ are
linearly dependent, but no $d-1$ columns are such. The row space
of the submatrix composed of these first $d$ columns has dimension
$d-1$ and a sequence of length $d$ belongs to this row space if
and only if its weight is even. By elementary row operations, we
can obtain a parity-check matrix of the form
\begin{equation} \label{Eq_badH1}
{\bf H}=\left(\begin{array}{cc}
{\bf H}'_d & {\bf H}' \\
{\bf 0}   & {\bf H}''
\end{array}\right),
\end{equation}
for some matrices ${\bf H}'$ and ${\bf H}''$ of appropriate sizes,
where ${\bf H}'_d$ is the $(d-1)\times d$ matrix given by
\begin{equation} \label{Eq_badH2}
{\bf H}'_d=\left(\begin{array}{cccccccc}
1 & 1 & 0 & 0 & 0 & \cdots & 0 & 0 \\
0 & 1 & 1 & 0 & 0 & \cdots & 0 & 0 \\
1 & 1 & 1 & 1 & 0 & \cdots & 0 & 0 \\
0 & 0 & 0 & 1 & 1 & \cdots & 0 & 0 \\
\vdots & \vdots & \vdots & \vdots & \vdots & \ddots & \vdots & \vdots \\
0 & 0 & 0 & 0 & 0 & \cdots & 1 & 1
\end{array}\right).
\end{equation}
Clearly, ${\cal S}=\{1,2,3\}$ is a stopping set for $\bf H$ as no
row of ${\bf H}_{\cal S}$ has weight one. \endofproof

Contrary to the weight enumerator and the incorrigible set
enumerator, which are fixed for a code $\cal C$, the stopping and
dead-end set enumerators depend on the choice of the parity-check
matrix $\bf H$. Theorem~\ref{badH} shows that no matter how large
the Hamming distance of the code is, a bad choice of the
parity-check matrix may lead to very poor performance. Therefore,
it is important to properly select the parity-check matrix of a
code when applying iterative decoding.

Clearly, adding rows to a parity-check matrix does not increase
any coefficient of the stopping set enumerator or the dead-end set
enumerator. On the contrary, these coefficients may actually
decrease at the expense of higher decoding complexity. The rows to
be added should be in the dual code ${\cal C}^{\perp}$ of $\cal
C$. By having all $2^{n-k}$ codewords in ${\cal C}^{\perp}$ as
rows, we obtain a parity-check matrix that gives the best possible
performance, but also the highest complexity, when applying
iterative decoding. Since the order of the rows does not affect
the decoding result, we refer to such matrix, with some ordering
imposed on its rows which is irrelevant to our work, as the
complete parity-check matrix of the code $\cal C$, and denote it
by ${\bf H}^\star$. Its stopping set enumerator is denoted by
$S^\star(x)=\sum_{i=0}^n S^\star_i x^i$, its dead-end set
enumerator by $D^\star(x)=\sum_{i=0}^n D^\star_i x^i$, and its
stopping distance by $s^\star$. Since the support of any codeword
is a stopping set for any parity-check matrix, we have
\begin{equation} \label{si}
\mbox{$S_i\ge S^\star_i\ge A_i$ and $D_i\ge D^\star_i\ge I_i$
$\forall i=0,1,\ldots,n$.}
\end{equation}
Consequently, $s\le s^\star \le d$, and $S^\star(x)$ and
$D^\star(x)$ are called the code's optimal stopping set enumerator
and optimal dead-end set enumerator, respectively. Schwartz and
Vardy \cite{SV} have shown that
\begin{equation}s^\star=d\end{equation} and the results derived
recently by Hollmann and Tolhuizen \cite{HT} imply, in addition,
that \begin{equation}S^\star_d=A_d\end{equation} and
\begin{equation}\label{dstar}D^\star(x)=I(x).\end{equation}

Actually, Schwartz and Vardy \cite{SV} have shown that,
for $d\ge 3$, it is
possible to construct a parity-check matrix with at most
$\sum_{i=1}^{d-2} \left({n-k\atop i}\right)$ rows for which $s=d$.
They also obtain interesting results on the minimum number of rows
in a parity-check matrix for which $s=d$. They obtain general
bounds on this minimum number, which they call the stopping
redundancy, as well as bounds for specific codes such as the Golay
code and Reed-Muller codes. Han and Siegel \cite{HS} derived
another general upper bound on the stopping redundancy for $d\ge 2$
given by $\sum_{i=1}^{\lceil(d-1)/2\rceil} \left({n-k\atop 2i-1}\right)$.

Hollmann and Tolhuizen \cite{HT} specified rows that can be formed
from any $(n-k)\times n$ parity-check matrix of rank $n-k$ to
yield a parity-check matrix for which $D_i=I_i$ for $0\le i\le m$,
where $m$ is any given integer such that $2\le m\le n-k$. They
have shown that the number of rows in the smallest parity-check
matrix achieving this is at most $\sum_{i=0}^{m-1}
\left({n-k-1\atop i}\right).$

\begin{example} \label{Ex1}
Let $\cal C$ be the $[8,4,4]$ Reed-Muller code. One of the
parity-check matrices of $\cal C$ is
\begin{equation} \label{Eq_H7}
\renewcommand\arraystretch{0.9}
{\bf H}_8=\left(\begin{array}{cccccccc}
1 & 0 & 1 & 0 & 1 & 0 & 1 & 0 \\
0 & 1 & 0 & 1 & 0 & 1 & 0 & 1 \\
0 & 0 & 1 & 1 & 0 & 0 & 1 & 1 \\
0 & 0 & 0 & 0 & 1 & 1 & 1 & 1 \\
1 & 1 & 1 & 1 & 0 & 0 & 0 & 0 \\
1 & 1 & 0 & 0 & 1 & 1 & 0 & 0 \\
0 & 1 & 1 & 0 & 1 & 0 & 0 & 1 \\
1 & 0 & 0 & 1 & 0 & 1 & 1 & 0 \\
\end{array}\right).
\end{equation}
For $i=4,5,6,7$, deleting the last $8-i$ rows in ${\bf H}_8$ still
gives a parity-check matrix ${\bf H}_i$ for the code $\cal C$.
Table~I gives the stopping set enumerator $S(x)$ and the dead-end
set enumerator $D(x)$ for the parity-check matrix ${\bf H}_i$ for
$i=4$, $5$, and $8$. The table also gives $S^\star(x)$ and
$D^\star(x)$ corresponding to the complete parity-check matrix
${\bf H}^\star$, and the weight enumerator $A(x)$ and the
incorrigible set enumerator $I(x)$. We point out that the matrix
${\bf H}_4$ is a frequently used full rank matrix for this code.
For this matrix $s=3$. The matrix ${\bf H}_5$ is the matrix
proposed by Schwartz and Vardy \cite{SV} to achieve $s=4$. For
this matrix $S_4>A_4$. The matrix ${\bf H}_8$ is constructed based
on the techniques proposed by Hollmann and Tolhuizen to achieve
$S_4=A_4$. For later purposes, we also define the matrix ${\bf
H}_{14}$ whose rows are the fourteen non-zero codewords in ${\cal
C}^{\perp}={\cal C}$ of weight four. The stopping set enumerator
and the dead-end set enumerator for this parity-check matrix are
also listed in the table.
\end{example}

\begin{table*}
\renewcommand\arraystretch{1.4}
\caption{$A(x)$ and $I(x)$ for the $[8,4,4]$ Reed-Muller code and
$S(x)$ and $D(x)$ for the parity-check matrices ${\bf H}_4$, ${\bf
H}_5$, ${\bf H}_8$, ${\bf H}_{14}$, and ${\bf H}^\star$.}
$$\begin{array}{|c|c|c|} \hline
& A(x) &I(x) \\ \hline & 1+14x^{4}+x^8 & 14x^4+56x^5+28x^6+8x^7+x^8 \\
\hline \hline \mbox{parity-check matrix} & S(x) & D(x) \\
\hline{\bf H}_4 & 1+2x^3+24x^4+40x^5+28x^6+8x^7+x^8 &
2x^3+32x^4+56x^5+28x^6+8x^7+x^8 \\ \hline {\bf H}_5 &
1+18x^4+36x^5+28x^6+8x^7+x^8 & 18x^4+56x^5+28x^6+8x^7+x^8 \\
\hline {\bf H}_8 & 1+14x^4+24x^5+28x^6+8x^7+x^8 &
14x^4+56x^5+28x^6+8x^7+x^8 \\ \hline {\bf H}_{14} &
1+14x^4+28x^6+8x^7+x^8 & 14x^4+56x^5+28x^6+8x^7+x^8
\\ \hline {\bf H}^\star &
1+14x^4+28x^6+8x^7+x^8 & 14x^4+56x^5+28x^6+8x^7+x^8 \\
\hline
\end{array}$$
\end{table*}

\section{Dead-End Set Results} \label{sec ds}

In this section, we investigate parity-check matrices for which
the iterative decoding procedure achieves optimal performance,
i.e., for which
\begin{equation} \label{sisdcor}
P^{\rm IT}_{\rm UD}({\bf H})=P^{\rm OPT}_{\rm UD}({\cal C}).
\end{equation}
In order to satisfy (\ref{sisdcor}), it is necessary and
sufficient that the dead-end set enumerator equals the
incorrigible set enumerator, i.e., $D(x)=I(x)$. From
(\ref{dstar}), we know that this is the case for the complete
parity-check matrix, which contains $2^{n-k}$ rows. However, from
a decoding complexity point of view, it may be desirable or
required to reduce the number of rows in the parity-check matrix.
Hence, an interesting research challenge is to find a parity check
matrix $\bf H$ for code $\cal C$, with a minimum number of rows,
but still having $D(x)=I(x)$.

As stated before, it is shown in \cite{HT} that there exists a
parity-check matrix $\bf H$ with at most $\sum_{i=0}^{m-1}
\left({n-k-1\atop i}\right)$ rows for which $D_i=I_i$, $0\le i\le
m$, for any $1\le m\le n-k$. By taking $m=n-k$ and noticing that
$D_i=I_i$ for $n-k+1\le i\le n$ from (\ref{basicI}) and
(\ref{basicD}), we deduce the following result.
\begin{theorem}[Hollmann and Tolhuizen] \label{holtol}
Let $\cal C$ be an $[n,k,d]$ binary linear code with $k<n$. Then,
there exists a parity-check matrix with at most $2^{n-k-1}$ rows
for which $D(x)=I(x)$.
\end{theorem}

Hollmann and Tolhuizen also show that for some codes, and in
particular for Hamming codes, $D(x)\not=I(x)$ for any parity-check
matrix with less than $2^{n-k-1}$ rows. However, depending on the
code, it may be possible to reduce the number of rows in a
parity-check matrix for which $D(x)=I(x)$ below $2^{n-k-1}$ as we
show next.

\begin{theorem} \label{di1}
Let $\bf H$ be the matrix whose rows are the non-zero codewords in
${\cal C}^\perp$ of weight at most $k+1$. Then, $\bf H$ is a
parity-check matrix for $\cal C$ and for this matrix $D(x)=I(x)$.
\end{theorem}

\beginofproof
Let ${\bf H}'$ be an $(n-k)\times n$ parity-check matrix for the
code $\cal C$. Then, there is a subset ${\cal S}$ of
$\{1,2,\ldots,n\}$ of size $n-k$ such that ${\bf H}'_{\cal S}$ is
an $(n-k)\times(n-k)$ matrix of rank $n-k$. The row space of this
matrix contains every unit weight vector of length $n-k$.
Therefore, the row space of ${\bf H}'$ contains $n-k$ vectors such
that each vector has exactly a single one in a unique position
indexed by an element in ${\cal S}$. Since these vectors have
weight at most $k+1$ and are linearly independent, it follows that
$\bf H$, which contains all of them as rows, has rank $n-k$ and is
indeed a parity-check matrix for $\cal C$.

Next, we prove that for this matrix $D(x)=I(x)$, i.e., $D_i=I_i$
for $i=0,1,\ldots,n$. From (\ref{012}), (\ref{basicI}), and
(\ref{basicD}), it suffices to show that $D_i=I_i$ for $3\le i\le
n-k$. For such an $i$, assume that ${\cal S}'$ is a subset of
$\{1,2,\ldots,n\}$ of size $i$ which does not contain the support
of a non-zero codeword. Then, the columns of the $(n-k)\times n$
parity-check matrix ${\bf H}'$ indexed by the elements in ${\cal
S}'$ are linearly independent. As ${\bf H}'$ has rank $n-k$, there
is a set ${\cal S}''$ such that ${\cal S}'\subseteq{\cal
S}''\subseteq\{1,2,\ldots,n\}$ and ${\bf H}'_{{\cal S}''}$ is an
$(n-k)\times(n-k)$ matrix of rank $n-k$. From the argument given
in the first part of this proof, ${\bf H}$ contains $n-k$ vectors
such that each vector has exactly a single one in a unique
position indexed by an element in ${\cal S}''$, and in particular
each vector has weight at most $k+1$. The existence of any one of
the $i$ vectors with a single one in a position indexed by an
element in ${\cal S}'$ proves that ${\cal S}'$ is not a stopping
set for ${\bf H}$. We conclude that every stopping set of size $i$
for $\bf H$ contains the support of a non-zero codeword. Hence,
$D_i=I_i$ for all $i$. \endofproof

Let $H(x)$ denote the well-known binary entropy function $-x\log_2
x-(1-x)\log_2(1-x)$ for $0<x<1$.
%which is extended to $0\le x\le
%1$ by defining $H(0)=H(1)=0$.

\begin{theorem} \label{dopt}
Let $\cal C$ be an $[n,k,d]$ binary linear code with $k\le n/2-1$.
Then, there exists a parity-check matrix with at most
$2^{nH((k+1)/n)}$ rows for which $D(x)=I(x)$.
\end{theorem}

\beginofproof
From the bounds on the sum of binomial coefficients as presented
on page 310 of \cite{MS}, it follows that the number of codewords
in the dual code of weight less than or equal to $k+1$ is at most
equal to $2^{nH((k+1)/n)}$. Hence, the result follows from
Theorem~\ref{di1}. \endofproof

Note that the bound from Theorem~\ref{dopt} improves upon the
bound from Theorem~\ref{holtol} for low-rate codes.
%which is illustrated in Figure~\ref{bounds}.
%\begin{corollary} \label{cordopt}
%Let $\cal C$ be an $[n,k,d]$ binary linear code with $k<n$. Then,
%there exists a parity-check matrix with at most $2^{0.773n}$ rows
%for which $D(x)=I(x)$.
%\end{corollary}

%\begin{figure}
%\includegraphics[width=3.45in]{deadendbounds.eps}
%\caption{Lower bounds (LB, dashed curves) and upper bounds (UB,
%solid curves) on $\psi^\ast$ as a function of the code rate
%$R=k/n$ for codes of lengths $n=20$ and $n=100$.}  \label{bounds}
%\end{figure}

\section{Stopping Set Results} \label{sec ss}

As stated earlier, iterative decoding based on a parity-check
matrix is optimal, in the sense of having the smallest possible
unsuccessful decoding probability on the binary erasure channel,
if and only if $D(x)$ for this matrix is identical to $I(x)$ for
code $\cal C$. This holds for the complete parity-check matrix as
well as other matrices, whose sizes are bounded in
Section~\ref{sec ds}. For $D(x)$ to be identical to $I(x)$, we
should have $s=d$ and $S_d=A_d$. Table~I shows that it is possible
to achieve optimal decoding using parity-check matrices, such as
${\bf H}_8$, with much smaller number of rows than in the complete
parity-check matrix ${\bf H}^\star$. This is true in spite of the
fact that these smaller matrices have stopping set enumerators
that are different from $S^\star(x)$. We may wonder then what is
the effect, if any, of the stopping set coefficients $S_i$ for
$i>d$ on performance. Notice that in this paper we defined the
probability of unsuccessful decoding as the probability that the
decoder fails to retrieve the transmitted codeword. Although an
iterative decoder is unsuccessful in case the erasure set is a
dead-end set, it still succeeds in retrieving those erased bits
whose indices do not belong to any of the stopping sets contained
in the erasure set. Therefore, it may be desirable to choose
parity-check matrices for which $S_i=A_i$ not only for $i=d$ but
also for $i>d$. Since $S^\star(x)\not=A(x)$ in Example~\ref{Ex1},
it follows that this is not possible in general. In fact,
Theorem~\ref{ThmSstarA} will show that $S^\star(x)=A(x)$ only for
a rather degenerate class of codes. Hence, the best that we may
hope for is to have parity-check matrices, smaller than the
complete parity-check matrix to reduce complexity, for which
$S(x)=S^\star(x)$. The matrix ${\bf H}_{14}$, specified in
Theorem~\ref{di1}, is one such matrix for the $[8,4,4]$
Reed-Muller code. Actually, it can be checked that this is the
smallest parity-check matrix for this code satisfying
$S(x)=S^\star(x)$.

\begin{example} \label{Ex2}
Let ${\cal C}$ be the $[8,4,4]$ Reed-Muller code considered in
Example~\ref{Ex1}. From Table~I, we notice that the iterative
decoders based on ${\bf H}^\star$, ${\bf H}_{14}$, and ${\bf H}_8$
achieve the smallest possible probability of unsuccessful
decoding, while the iterative decoders based on ${\bf H}_4$ and
${\bf H}_5$ do not. Although ${\bf H}_{14}$ is larger than ${\bf
H}_8$ and both achieve the maximum successful decoding
probability, there are advantages in using ${\bf H}_{14}$ instead
of ${\bf H}_8$. For instance, suppose that the erasure set is
$\{1,2,3,7,8\}$. This erasure set is an incorrigible set since it
contains $\{1,2,7,8\}$ which is the support of a non-zero codeword
in the code. Therefore, any decoding method fails in retrieving
the transmitted codeword. However, iterative decoding based on
${\bf H}_{14}$ succeeds in determining the erased bit $c_3$ from
the parity-check equation $c_3+c_4+c_5+c_6=0$ since $(00111100)$
is a row in ${\bf H}_{14}$. On the other hand, iterative decoding
based on ${\bf H}_8$ does not succeed in retrieving any of the
erased bits. Actually, since the coefficient of $x^5$ in the
stopping set enumerator of ${\bf H}_{14}$ is zero, it follows that
if the erasure set is a dead-end set of size five, it is always
possible to retrieve one of the erased bits using ${\bf H}_{14}$.
This is not true if matrix ${\bf H}_8$ is used instead.
\end{example}

We will show that the range $0\le i\le 2$ specified by (\ref{012})
for which $S_i=A_i$ can be considerably extended in case $\bf H$
is the complete parity-check matrix ${\bf H}^\star$. First we
start with two lemmas.
\begin{lemma} \label{ldi0}
For any non-empty set ${\cal S}\subseteq \{1,2,\ldots,n\}$ which
does not contain the support of a non-zero codeword, each binary
vector of length $|{\cal S}|$ appears exactly $2^{n-k-|{\cal S}|}$
times as a row in ${\bf H}^\star_{\cal S}$.
\end{lemma}

\beginofproof Since $\cal S$ does not contain the support of a non-zero codeword,
the $2^{n-k}\times |{\cal S}|$ matrix ${\bf H}^\star_{\cal S}$ has
rank $|{\cal S}|$. Hence, there are $|{\cal S}|$ linearly
independent rows in ${\bf H}^\star_{\cal S}$. The linear
combinations of these $|{\cal S}|$ rows generate the space of all
binary vectors of length $|{\cal S}|$, and thus each of these
$2^{|{\cal S}|}$ vectors appears exactly $2^{n-k}/2^{|{\cal S}|}$
times as a row in ${\bf H}^\star_{\cal S}$. \endofproof

\begin{lemma} \label{ldi1}
For any set ${\cal S}\subseteq \{1,2,\ldots,n\}$ which contains
exactly one support ${\cal S}'$ of a non-zero codeword, each
binary vector of length $|{\cal S}|$, with even weight on the
positions indexed by ${\cal S}'$ and any weight on the positions
indexed by ${\cal S}\setminus{\cal S}'$, appears exactly
$2^{n-k-|{\cal S}|+1}$ times as a row in ${\bf H}^\star_{\cal S}$.
\end{lemma}

\beginofproof Since $\cal S$ contains exactly one support ${\cal S}'$ of
a non-zero codeword, the $2^{n-k}\times |{\cal S}|$ matrix ${\bf
H}^\star_{\cal S}$ has rank $|{\cal S}|-1$. Hence, there are
$|{\cal S}|-1$ linearly independent rows in ${\bf H}^\star_{\cal
S}$. The linear combinations of these $|{\cal S}|-1$ rows generate
the space of all binary vectors of length $|{\cal S}|$ with even
weight on the positions indexed by ${\cal S}'$ and any weight on
the positions indexed by ${\cal S}\setminus{\cal S}'$, and thus
each of these $2^{|{\cal S}|-1}$ vectors appears exactly
$2^{n-k}/2^{|{\cal S}|-1}$ times as a row in ${\bf H}^\star_{\cal
S}$. \endofproof

\begin{theorem} \label{shs}
For any code,
\begin{equation} \label{sis}
S_i^\star  =  A_i  \mbox{ for $i=0,1,\ldots,\min\{\lceil
3d/2\rceil-1,n\}$,}
\end{equation}
i.e., the enumerators $S^\star(x)$ and $A(x)$ are equal in at
least the first $\min\{\lceil 3d/2\rceil,n+1\}$ coefficients.
\end{theorem}

\beginofproof Since the result is trivial for $i=0$, we may assume
$1\le i\le\min\{\lceil 3d/2\rceil-1,n\}$. Suppose that $\cal S$ is
a stopping set of size $i$ for ${\bf H}^\star$, which is not the
support of a codeword. This set $\cal S$ contains at most one
support of a non-zero codeword, since it follows from the Griesmer
bound \cite{MS} that any linear code of dimension greater than $1$
and Hamming distance at least $d$ has a length of at least
$d+\lceil d/2\rceil=\lceil 3d/2\rceil$. It follows from
Lemmas~\ref{ldi0} and \ref{ldi1} that ${\bf H}^\star_{\cal S}$
contains at least one row of weight one. Together with
(\ref{eqSS}), we reach a contradiction to the assumption that
$\cal S$ is a stopping set for ${\bf H}^\star$. Hence, any
stopping set of size $i$ for ${\bf H}^\star$ is the support of a
codeword. In conclusion, $S_i^\star \le A_i$, and together with
(\ref{si}) we obtain the result presented in (\ref{sis}).
\endofproof

%Based on (\ref{basicS} and the proof of Theorem~\ref{di1}, we can
%also derive the following result

%\begin{theorem} \label{di2}
%Let $\bf H$ be the matrix whose rows are the non-zero codewords in
%${\cal C}^\perp$ of weight at most $k+1$. Then, $\bf H$ is a
%parity-check matrix for $\cal C$ and for this matrix
%\begin{equation}S^\star_i=A_i \mbox{ if $0\le i\le n-k$}.\end{equation}
%\end{theorem}

In the remainder of this section, we give a complete
characterization of codes that have parity-check matrices for
which $S(x)=A(x)$, i.e., codes with parity-check matrices for
which every stopping set is a support of a codeword. For
convenience, such codes are called minimum stopping. From
(\ref{si}), we conclude that a code is minimum stopping if and
only if its optimal stopping set enumerator equals its weight
enumerator, i.e., $S^\star(x)=A(x)$. We start by giving three
classes of codes satisfying this condition.

%In the remainder of this section, we give a complete
%characterization of codes for which $S^\star(x)=A(x)$, i.e., codes
%satisfying the condition that every stopping set for their
%complete parity-check matrices is the support of a codeword. We
%start by giving three classes of codes satisfying this condition.

(i) ${\cal R}_n$ is the $[n,1,n]$ repetition code consisting of
the all-zero and all-one vectors of length $n$. From
Theorem~\ref{shs}, it follows that $S_i^\star=A_i$ for
$i=0,1,\ldots,n$. Hence, $S_0^\star=A_0=1$, $S_n^\star=A_n=1$, and
$S_i^\star=A_i=0$ for $i=1,2,\ldots,n-1$.

(ii) ${\cal F}_n$ is the $[n,n,1]$ full-code consisting of all
binary vectors of length $n$. Clearly,
$S_i^\star=A_i=\left({n\atop i}\right)$ for $i=0,1,\ldots,n$.

(iii) ${\cal Z}_n$ is the $[n,0,\infty]$ zero-code consisting of
one codeword only, which is the all-zero vector of length $n$.
Since all vectors of length $n$, including those of weight one,
belong to the complete parity-check matrix of the code, it follows
that $S_i^\star=A_i=0$ for $i=1,2,\ldots,n$, and
$S_0^\star=A_0=1$.

Next, we introduce a useful notation. For $i=1,2,\ldots,t$, let
${\cal C}_i$ be a binary linear $[n_i,k_i,d_i]$ block code. Then,
we define
%\begin{equation}  \label{Eq_def1}
\begin{eqnarray}
%\begin{eqnarray*}
{\cal C}_1\oplus{\cal C}_2\oplus\cdots\oplus{\cal C}_t&=&\{({\bf
c}_1,{\bf c}_2,\ldots,{\bf c}_t):
   {\bf c}_i\in{\cal C}_i  \nonumber \\
   & &     \;\;\;\;\;\forall i=1,2,\ldots,t\}, \label{CCt}
\end{eqnarray}
%\end{equation}
i.e., ${\cal C}_1\oplus{\cal C}_2\oplus\cdots\oplus{\cal C}_t$ is
the set of all sequences obtained by juxtaposing codewords in
${\cal C}_1,{\cal C}_2,\ldots,{\cal C}_t$ in this order. Clearly,
${\cal C}_1\oplus{\cal C}_2\oplus\cdots\oplus{\cal C}_t$ is an
$[n_1+n_2+\cdots+n_t, k_1+k_2+\cdots+k_t,
\min\{d_1,d_2,\ldots,d_t\}]$ binary linear block code.

Finally, recall that two codes are equivalent if there is a fixed
permutation of indices that maps one to the other \cite{MS}, and
that a code is said to have no zero-coordinates if and only if
there is no index $i\in\{1,2,\ldots,n\}$ such that $c_i=0$ for
every codeword $(c_1,c_2,\ldots,c_n)$.
%Clearly, if one of the codes has a stopping set which is the
%support of a codeword, then the other code also does.

\begin{lemma} \label{lem CCCt}
The code ${\cal C}_1\oplus{\cal C}_2\oplus\cdots\oplus{\cal C}_t$
is minimum stopping if and only if ${\cal C}_i$ is minimum
stopping for all $i=1,\ldots,t$.
%The code ${\cal C}_1\oplus{\cal C}_2\oplus\cdots\oplus{\cal C}_t$
%has optimal stopping set enumerator if and only if ${\cal C}_i$
%has optimal stopping set enumerator for all $i=1,\ldots,t$.
\end{lemma}
\beginofproof
First, notice that the code ${\cal C}={\cal C}_1\oplus{\cal
C}_2\oplus\cdots\oplus{\cal C}_t$ has a block diagonal parity
check matrix ${\bf H}_1\oplus{\bf H}_2\oplus\cdots\oplus{\bf H}_t$
defined by
\begin{equation} \label{HoH}
{\bf H}_1\oplus{\bf H}_2\oplus\cdots\oplus{\bf
H}_t=\left(\begin{array}{cccc}
{\bf H}_1 & {\bf 0} & \cdots & {\bf 0} \\
{\bf 0} & {\bf H}_2 & \cdots & {\bf 0} \\
\vdots  & \vdots    & \ddots & \vdots  \\
{\bf 0} & {\bf 0}   & \cdots &{\bf H}_t
\end{array}\right),
\end{equation}
where ${\bf H}_i$ is a parity check matrix for ${\cal C}_i$. The
complete parity-check matrix ${\bf H}^\star$ of $\cal C$ has all
elements of the row space of the matrix from (\ref{HoH}) as its
rows.

Next, let $\cal S$ be a subset of
$\{1,2,\ldots,n_1+n_2+\cdots+n_t\}$. For $i=1,2,\ldots,t$, define
\begin{equation}
{\cal S}_i=\{1\le m\le n_i: m+\sum_{j=1}^{i-1} n_j \in {\cal S}\}.
\end{equation} Then, it follows from (\ref{CCt})
that $\cal S$ is the support of a codeword in ${\cal
C}_1\oplus{\cal C}_2\oplus\cdots\oplus{\cal C}_t$ if and only if
${\cal S}_i$ is the support of a codeword in ${\cal C}_i$ for all
$i=1,2,\ldots,t$.

Further, ${\cal S}$ is a stopping set for ${\bf H}^\star$ if and
only if ${\cal S}_i$ is a stopping set for ${\bf H}^\star_i$ for
all $i=1,2,\ldots,t$. This follows from the fact that a sequence
${\bf r}$ is a row in ${\bf H}^\star$ if and only if it can be
written as ${\bf r}=({\bf r}_1,{\bf r}_2,\ldots,{\bf r}_t)$ where
${\bf r}_i$ is a row in ${\bf H}^\star_i$ for all
$i=1,2,\ldots,t$. Hence, a sequence ${\bf r}$ is a row in ${\bf H}
^\star_{\cal S}$ if and only if it can be written as ${\bf
r}=({\bf r}_1,{\bf r}_2,\ldots,{\bf r}_t)$ where ${\bf r}_i$ is a
row in ${\bf H}^\star_{i,{\cal S}_i}$ for all $i=1,2,\ldots,t$. If
for some $i=1,2,\ldots,t$, ${\cal S}_i$ is not a stopping set for
${\bf H}^\star_i$, then ${\bf H}^\star_{i,{\cal S}_i}$ has a row
of weight one. Juxtaposing this row with the all-zero rows in
${\bf H}^\star_{j,{\cal S}_j}$, for all $j\not=i$, gives a row in
${\bf H}^\star_{\cal S}$ of weight one. This implies that ${\cal
S}$ is not a stopping set for ${\bf H}^\star$. On the other hand,
if for all $i=1,2,\ldots,t$, ${\cal S}_i$ is a stopping set for
${\bf H}^\star_i$, then for all $i$, ${\bf H}^\star_{i,{\cal
S}_i}$ has no row of weight one. Juxtaposing rows of weights other
than one yields a row of weight other than one. Hence, $\cal S$ is
a stopping set for ${\bf H}^\star$.

We conclude that $\cal S$ is a stopping set for ${\bf H}^\star$
which is the support of a codeword in ${\cal C}_1\oplus{\cal
C}_2\oplus\cdots\oplus{\cal C}_t$ if and only if, for all
$i=1,2,\ldots,t$, ${\cal S}_i$ is a stopping set for ${\bf
H}_i^\star$ which is the support of a codeword in ${\cal C}_i$.
Hence, ${\cal C}_1\oplus{\cal C}_2\oplus\cdots\oplus{\cal C}_t$ is
minimum stopping if and only if, for all $i=1,2,\ldots,t$, ${\cal
C}_i$ is minimum stopping.
%C}_t$ does not have optimal stopping set enumerator if and only
%if, for some $i=1,2,\ldots,t$, ${\cal C}_i$ does not have optimal
%stopping set enumerator.
\endofproof

\begin{lemma} \label{Lem_4}
Let $\cal C$ be a minimum stopping binary linear $[n,k,d]$ block
code with $d\ge 2$ and no zero-coordinates.
%Let $\cal C$ be a binary linear $[n,k,d]$ block code with $d\ge 2$
%and no zero-coordinates which has optimal stopping set enumerator.
If $d=n$, then ${\cal C}$ is ${\cal R}_n$. Otherwise, $k\ge 2$ and
$\cal C$ is equivalent to ${\cal R}_d\oplus{\cal C}''$ for some
binary linear $[n-d,k-1,d'']$ block code ${\cal C}''$ with $d''\ge
2$ and no zero-coordinates.
\end{lemma}
\beginofproof
Up to equivalence, we may assume that $\cal C$ has a codeword
composed of $d$ ones followed by $n-d$ zeros. In particular, the
first $d$ columns in any given parity check matrix of $\cal C$ are
linearly dependent, but no $d-1$ columns are such. The row space
of the submatrix composed of these first $d$ columns has dimension
$d-1$ and a sequence of length $d$ belongs to this row space if
and only if its weight is even. Therefore, in case $d=n$, $\cal C$
has a full-rank $(d-1)\times d$ parity check matrix of the form
\begin{equation} \label{Eq_PCMR}
{\bf H}={\bf H}''_d=\left(\begin{array}{ccccc}
1 & 1 & 0 & \cdots & 0 \\
1 & 0 & 1 & \cdots & 0 \\
\vdots & \vdots &   & \ddots & \vdots \\
1 & 0 & 0 & \cdots & 1
\end{array}\right),
\end{equation}
which shows that $\cal C$ is the repetition code of length $n$.
Further, in case $d\le n-1$, $\cal C$ has a full-rank $(n-k)\times
n$ parity check matrix of the form
\begin{equation} \label{Eq_Lem31}
{\bf H}=\left(\begin{array}{cc}
{\bf H}''_d & {\bf H}' \\
{\bf 0}              & {\bf H}''
\end{array}\right),
\end{equation}
where ${\bf H}'$ and ${\bf H}''$ are $(d-1)\times (n-d)$ and
$(n-k-d+1)\times (n-d)$ matrices, respectively. Notice that ${\bf
H}''$ has at least one row since $d\le n-k+1$ by the Singleton
bound and if equality holds with $2\le d\le n-1$, then $\cal C$ is
the even weight code of length $n\ge 3$ which has $\{1,2,3\}$ as a
stopping set for ${\bf H}^\star$ which is not the support of a
codeword. This contradicts the assumption that $\cal C$ is minimum
stopping. Clearly, ${\bf H}''$ is a matrix
%codeword. This contradicts the assumption that $\cal C$ has
%optimal stopping set enumerator. Clearly, ${\bf H}''$ is a matrix
of rank $n-k-d+1$ since ${\bf H}$ in (\ref{Eq_Lem31}) is a
full-rank matrix. If $k=1$, then ${\bf H}''$ has rank $n-d$ and
${\cal C}$ has zero-coordinates. Therefore, $k\ge 2$. To complete
the proof, it suffices to show that the row space of ${\bf H}'$ is
a subspace of the row space of ${\bf H}''$ since, in this case, by
elementary row operations, $\cal C$ has a parity-check matrix
\begin{equation} \label{Eq_Lem31a}
{\bf H}=\left(\begin{array}{cc}
{\bf H}''_d & {\bf 0} \\
{\bf 0}              & {\bf H}''
\end{array}\right),
\end{equation} and thus
${\cal C}={\cal R}_d\oplus{\cal C}''$ where ${\cal C}''$ is the
code with parity check matrix ${\bf H}''$. This code has length
$n-d$, dimension $k-1$, and Hamming distance $d''\ge d\ge 2$ with
no zero-coordinates. Now, suppose, to get a contradiction, that
the above is not true, i.e., the row space of ${\bf H}'$ is not a
subspace of the row space of ${\bf H}''$. Then, the null space of
${\bf H}''$ is not a subspace of the null space of ${\bf H}'$. Let
${\bf c}''$ be a vector of length $n-d$ which belongs to the null
space of ${\bf H}''$ but not to the null space of ${\bf H}'$. Up
to equivalence, we may assume that ${\bf c}''$ is composed of $w$
ones followed by $n-d-w$ zeros, where $w$, $1\le w\le n-d$, is the
weight of ${\bf c}''$. We claim that $\{1,2,\ldots,d+w\}$ is a
stopping set for ${\bf H}^\star$ which is not the support of a
codeword in ${\cal C}$. From (\ref{Eq_Lem31}), we have
\begin{equation} \label{Eq_Lem32}
{\bf H}_{\{1,2,\ldots,d+w\}}=\left(\begin{array}{cc}
{\bf H}''_d & {\bf H}'_{\{1,2,\ldots,w\}} \\
{\bf 0}              & {\bf H}''_{\{1,2,\ldots,w\}}
\end{array}\right).
\end{equation}
Notice that any nontrivial linear combination of the rows of ${\bf
H}''_d$ yields a non-zero vector of even weight. Furthermore,
since ${\bf c}''$, which starts with $w$ ones followed by $n-d-w$
zeros, is in the null space of ${\bf H}''$, it follows that any
linear combination of the rows of ${\bf H}''_{\{1,2,\ldots,w\}}$
yields an even weight vector. We conclude that no linear
combination of the rows of ${\bf H}_{\{1,2,\ldots,d+w\}}$ yields a
vector of weight one. Hence, $\{1,2,\ldots,d+w\}$ is a stopping
set for ${\bf H}^\star$. Next, notice that if $\{1,2,\ldots,d+w\}$
is the support of a codeword in $\cal C$, then the columns in
${\bf H}_{\{1,2,\ldots,d+w\}}$ in (\ref{Eq_Lem32}) should add up
to zero. From (\ref{Eq_PCMR}), we know that the first $d$ columns
add up to zero. Therefore, the columns of ${\bf
H}'_{\{1,2,\ldots,w\}}$ should add up to the zero. However, this
cannot be the case as ${\bf c}''$, which starts with $w$ ones
followed by $n-d-w$ zeros, is not in the null space of ${\bf H}'$.
In conclusion, we have shown that $\{1,2,\ldots,d+w\}$ is a
stopping set for ${\bf H}^\star$ which is not the support of a
codeword in ${\cal C}$. This contradicts the fact that ${\cal C}$
is minimum stopping.
%has optimal stopping set enumerator.
\endofproof

\begin{lemma} \label{Lem_5}
If $\cal C$ is a minimum stopping binary linear $[n,k,d]$ block
code with $d\ge 2$ and no zero-coordinates,
%which has optimal stopping set enumerator,
then ${\cal C}$ is equivalent to \begin{equation}{\cal
R}_{n_1}\oplus{\cal R}_{n_2}\oplus\cdots\oplus{\cal
R}_{n_t},\end{equation} for some integers $n_1,n_2,\ldots,n_t\ge
2$ and $t\ge 1$ such that $n_1+n_2+\cdots+n_t=n$.
\end{lemma}
\beginofproof
The lemma trivially holds for all codes of lengths two or less. We
use induction and assume that it holds for all codes of length
less than $n$. From Lemma~\ref{Lem_4}, we know that either ${\cal
C}$ is equivalent to ${\cal R}_n$, which is consistent with the
statement of the lemma, or $k\ge 2$ and $\cal C$ is equivalent to
${\cal R}_d\oplus{\cal C}''$ for some binary linear
$[n-d,k-1,d'']$ block code ${\cal C}''$ with $d''\ge 2$ and no
zero-coordinates. From Lemma~\ref{lem CCCt}, we know that ${\cal
C}''$ is a minimum stopping code. Since ${\cal C}''$ has length
$n-d<n$, it follows from the induction hypothesis that ${\cal
C}''$ is equivalent to ${\cal R}_{m_1}\oplus\cdots\oplus{\cal
R}_{m_v}$, for some integers $m_1,\ldots,m_v\ge 2$ and $v\ge 1$
such that $m_1+\cdots+m_v=n-d$. Then, ${\cal R}_d\oplus{\cal C}''$
has the same form as given in the lemma.
\endofproof

\begin{theorem} \label{ThmSstarA}
A binary linear $[n,k,d]$ block code $\cal C$ is minimum stopping,
i.e., satisfies $S^\star(x)=A(x)$, if and only if it is equivalent
to
\begin{equation}{\cal R}_{n_1}\oplus{\cal R}_{n_2}\oplus\cdots\oplus{\cal
R}_{n_u}\oplus{\cal F}_{n_F}\oplus{\cal Z}_{n_Z},\end{equation}
for some nonnegative integers $n_1,n_2,\ldots,n_u,n_F,n_Z$ and
$u$, where $n_1,n_2,\ldots,n_u\ge 2$ and
$n_1+n_2+\cdots+n_u+n_F+n_Z=n$.
\end{theorem}

\beginofnote In the theorem, we allow $u=0$ in which case $\cal C$ is
equivalent to ${\cal F}_{n_F}\oplus{\cal Z}_{n_Z}$. We also allow
$n_F=0$ and/or $n_Z=0$, in which case the corresponding code with
length zero disappears from ${\cal R}_{n_1}\oplus{\cal
R}_{n_2}\oplus\cdots\oplus{\cal R}_{n_u}\oplus{\cal
F}_{n_F}\oplus{\cal Z}_{n_Z}$.

\beginofproof
The ``if''-part of the theorem follows from Lemma~\ref{lem CCCt}
and the observations that the $S^\star(x)=A(x)$ property holds for
any repetition code ${\cal R}_{n_i}$, the full-code ${\cal
F}_{n_F}$, and the zero-code ${\cal Z}_{n_Z}$.

Next, we proof the ``only if''-part of the theorem. Up to
equivalence, we may assume that ${\cal C}={\cal C}'\oplus{\cal
F}_{n_F}\oplus{\cal Z}_{n_Z}$, where $n_Z$ is the number of
zero-coordinates of $\cal C$, $n_F$ is the number of codewords of
weight one in $\cal C$, i.e., the number of all-zero columns in
any parity check matrix of $\cal C$, and ${\cal C}'$ is a binary
linear code of length $n-n_F-n_Z$ with $d\ge 2$ and no
zero-coordinates. Here we assume that if $n-n_F-n_Z$, $n_F$, or
$n_Z$ equal zero, then the corresponding code disappears from
${\cal C}'\oplus{\cal F}_{n_F}\oplus{\cal Z}_{n_Z}$. If ${\cal
C}'$ does not disappear, then it can be written as stated in
Lemma~\ref{Lem_5}.
\endofproof

\section{Conclusion} \label{Conc}

In this paper, we examined how the performance of iterative
decoding when applied to a binary linear block code over an
erasure channel depends on the parity-check matrix representing
the code. This code representation determines the complexity of
the decoder. We have shown that there is a trade-off between
performance and complexity.

In particular, we have shown that, regardless of the choice of the
parity-check matrix, the stopping set enumerator differs from the
weight enumerator except for a degenerate class of codes. In spite
of that, it is always possible to choose parity-check matrices for
which the dead-end set enumerator equals the incorrigible set
enumerator. Iterative decoding based on such matrices is optimal,
in the sense that it gives the same probability of unsuccessful
decoding on the binary erasure channel as an exhaustive decoder.
We presented bounds on the number of rows in parity-check matrices
with optimal dead-end set enumerators, thus bounding the
complexity of iterative decoding achieving optimal performance.

%\appendices

%\section{}

%\section*{Acknowledgment}


\end{document}